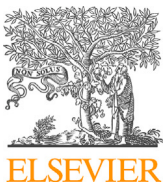
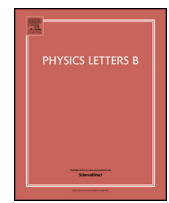

Letter

# Study of octupole deformations in Pb-Pb collisions at 5.02 TeV

Saraswati Pandey [a],[*], B.K. Singh [a,b]

[a] *Department of Physics, Institute of Science, Banaras Hindu University (BHU), Varanasi, 221005, India*
[b] *Discipline of Natural Sciences, PDPM Indian Institute of Information Technology Design & Manufacturing, Jabalpur-482005, India*

A R T I C L E  I N F O

Editor: M. Pierini

A B S T R A C T

In this letter, we present the study of the role of octupole deformation in non-spherical nuclei in most-central Pb-Pb collisions at LHC energy regime. The sensitivity of octupole deformation $\beta_3$ to the QGP observables is presented by employing Monte-Carlo HYDJET++ model. Motivated by the discrepancies in the $v_2$-to-$v_3$ puzzle found in Pb-Pb collisions and the low-energy nuclear structure calculations of nuclear deformation, we studied the first basic observables necessary for any study in heavy-ion collisions. Using HYDJET++ framework, we calculate the pseudorapidity distribution, transverse momentum ($p_T$) spectra and average anisotropic flow ($v_2$ and $v_3$) of primary charged hadrons with different parameters in two geometrical configurations: body-body and tip-tip type of Pb-Pb collisions. The kinematic ranges $0 < p_T < 20$ GeV/c and $|\eta| < 0.8$ are considered. We observe that the charged hadron multiplicity and transverse momentum spectra are dependent on the strength of octupole deformation parameter. $\langle v_2 \rangle$ and $\langle v_3 \rangle$ in body-body collisions show weak positive correlation with $\beta_3$ while average anisotropic flow in tip-tip collisions are weakly correlated to $\beta_3$ in most-central collision region.

(Ultra-) Relativistic heavy-ion collisions at BNL Relativistic Heavy Ion Collider (RHIC) and the CERN Large Hadon Collider (LHC) have been a means to explore the new state of matter, quark-gluon plasma (QGP) predicted by the theory of strong interactions, the Quantum Chromodynamics (QCD) [1]. It is predicted to be produced at the earlier stage of central nucleus-nucleus collisions. In the theoretical side of investigating QGP, the initial geometry fluctuations are studied by some models such as AMPT [2,3], while some methods also exist which have been proposed to study collective flow and geometry analysis in relation to the initial state fluctuations [4]. The collective flow and other QGP observables play important role in these experimental and theoretical investigations, and great progress has been made so far at relativistic energies [5].

Heavy-ion collisions phenomenology has reached a point in precision that the differences in the collective flow predictions between the beam energies differs only at the level of a few percent [6]. The various models vary in terms of the initial conditions and transport coefficients. However, it still is important to predict and study the basic and so called "bread and butter" observables such as charged particle multiplicity spectra, transverse momentum spectra, two-particle azimuthal anisotropies, etc [6–9]. In such a scenario, it is obvious to have deviations in experimental data from theoretical calculations in large systems which in turn signifies the missing physics. For instance, the recent measurement of two-particle elliptic flow showed an enhancement in $Xe^{129} - Xe^{129}$ central collisions compared to that assuming Xe to be a spherical nucleus [9]. The reason was associated to the small deformation ($\beta_2 = 0.162$ and $\beta_4 = -0.003$) existing in Xe-nucleus [10,11]. Similar reasons were responsible for the observed flow in ultra-central U-U collisions at RHIC energy regime [12,13]. So, typically, the most sensitive region to nuclear deformation are the most-central collisions. This is because, at zero impact parameter, the QGP signal for example, flow arises solely from fluctuations. In non-central collisions, the fluctuation sources scale with the collision system-size, and this having least effect in central collisions where the nuclear overlapping region is largest.

Being mindful of the recent studies on deformed nucleus-nucleus collisions, it is quite instinctive to question the nuclei who have studied considering them to be spherical in shape such as $Pb^{208}$. To surprise, Pb is a highly stable nucleus with doubly-magic atomic number. Tabulated nuclear ground-state masses and deformations from FRDM calculations state Pb nucleus to be perfectly spherical, with zero deformation [14]. However, tables from reduced transition probability studies suggest Pb to be deformed with positive values of quadrupole and octupole moments [15–17]. However, these nuclear structure studies have been performed at lower beam energies, several orders lower than relativistic heavy-ion collisions. Also, lower energy nuclear structure experiments involve electric charge density and its associated geometry while heavy-ion collisions probe the colour charge density and its associated geometry. This may cause a difference in the initial geometry in relevance to heavy-ion collisions. Also, the deformation parameters, $\beta_2$ and $\beta_3$ may differ between the electric and colour charge densities in the two scenarios. So, it is obligatory to test these deformation parameters in the domain of heavy-ion collisions.

The evidence of deformation in nuclei indicates that $Pb^{208}$ nucleus might exist in pear-like shape, having octupole deformation. However, evidences of quadrupole deformation also exist but the higher-order, the octupole moment is quite debative. In a recent study, considering Pb nucleus to be deformed, $v_2$-to-$v_3$ puzzle in ultra-central collisions was studied [6]. It was observed that the ratio between $v_2$ and $v_3$ slightly improved when Pb nucleus was considered to have significant octupole deformation, while the triangular cumulant ratio $v_3\{4\}/v_3\{2\}$ was found to get worse. This implied that the triangular cumulant ratio preferred no octupole deformation, with $\beta_3 \lesssim 0.0375$ for $Pb^{208}$, leaving the $v_2$-to-$v_3$ puzzle a challenge for hydrodynamic models. Thus, it is necessary to first perform a proper study of the role of octupole deformation on the basic QGP observables before working on the more complicated ones such as $v_2$-to-$v_3$ puzzle. The main motivation here is to study the effect of octupole deformation on the basic observables: pseudorapidity density spectra, transverse momentum distribution, and anisotropic flow ($v_2$ and $v_3$) spectra in Pb-Pb collisions at 5.02 TeV.

In the recent years, several studies have been performed on deformed collision systems [18–20]. A nucleus being deformed offers numerous geometries in which it can be collided. The two most pronounced ones are the body-type and the tip-type, depending upon the orientation of the nuclei. In this letter, we study the possible octupole deformation in Pb-Pb collisions at 5.02 TeV with the aim to see the effect of nuclear octupole deformation on the basic observables of QGP via heavy-ion collisions. In the presented work, we try to study the pseudorapidity density distribution, transverse momentum spectra, and average anisotropic flow harmonics of primary charged hadrons as a function of nuclear octupole deformation. For this purpose, we exploit HYDJET++ (HYDrodynamics plus JETs) which is a Monte-Carlo model designed to simulate (ultra-) relativistic heavy-ion collisions. It involves the superposition of two independent components: the soft hydro-type state and the hard state resulting from the medium-modified multi-parton fragmentation, simulating both the states simultaneously. The details on the physics of the model and its simulation procedure can be seen in the corresponding articles [21,22]. The main details of the model are as follows:-

The model for the hard state multi-parton production in a HYDJET++ event is based on Pythia Quenched (PYQUEN) model [23]. This involves the production of initial parton spectra in accordance to PYTHIA and generated vertices are measured at a given impact parameter. Basically, PYQUEN model changes a jet event produced by PYTHIA by generating nucleonic collision vertices in accordance with the Glauber model at a certain impact parameter. Then, a rescattering-by-rescattering simulation using an algorithm of the parton path in the dense zone and the associated radiative and collision energy losses takes place [24–28]. This is followed by the hadronization according to the Lund String Model [29] for hard partons and in medium emitted gluons. An important cold nuclear matter effect (shadowing of parton's distribution function) is included using an impact parameter-dependent parameterization under the framework of Glauber-Gribov theory [30,31]. In HYDJET++ model, the collisional energy loss due to scattering with low momentum transfer is not considered because of its little contribution to the total collisional energy loss in comparison to that with high momentum scattering [32,33]. The medium in which partonic rescattering takes place is boost invariant and the quark-gluon fluid expands in the longitudinal direction, and the produced partons lie on a hyper-surface of equal proper times $\tau$ [24].

The model for the soft state of HYDJET++ is a thermal hadronic state produced on the chemical and thermal freezeout hypersurfaces obtained from a parameterization of relativistic hydrodynamics with preset freezeout conditions [34,35]. In HYDJET++, it is assumed that the hadronic matter produced in a nuclear collision achieves local equilibrium after a short period of time (<1 fm/c) and then expands hydrodynamically. Here, the presumption of single freeze-out is eliminated as the particle densities at the chemical freezeout stage are too high to consider the particles as free streaming [36]. So, a more different scenario of differential freeze-out (thermal and chemical freeze-outs ($T_{ch} \geq T_{th}$)) is considered. The system is expected to go through a hydrodynamic expansion with frozen chemical composition in between these two freeze-outs followed by cooling and then the hadrons stream freely as the thermal freeze-out temperature is achieved.

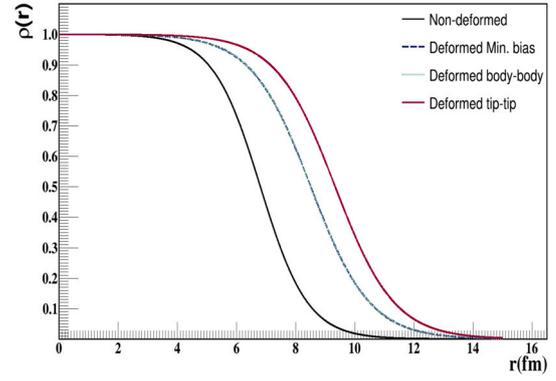

**Fig. 1.** The nuclear density profile for lead nucleus purely having octupole deformation $\beta_3 = 0.110$. Shown are the non-deformed and deformed Woods-Saxon nuclear density profiles. Two types of geometries are shown: body-type and tip-type of geometrical configuration [19].

The main modification made in the present (latest) version of HYDJET++ model is to modify the nuclear density profile function. Upon modification, the Woods-Saxon profile function in spherical coordinate system for deformed lead nucleus is expressed as:

$$\rho(r,z,\theta) = \frac{\rho_0}{1+\exp\left\{\frac{(r-R_A)}{a}\right\}} \quad (1)$$

where,
$\rho_0 = \rho_0^{const} + \text{correction}, \rho_0^{const} = \frac{M}{V} = \frac{3A}{4\pi R_A^3}$,
$R_A = R_0 A^{1/3}(1 + \beta_2 Y_{20} + \beta_3 Y_{30} + \beta_4 Y_{40})$,
where $R_0 = 1.15$ fm,
The correction term is calculated as= $\rho_0^{const} \not\supseteq (\pi f/R_A)^2$,
where $f = 0.54$ fm,
$\beta_n$ ($n = 2, 3, 4$) are the deformation parameters,
a=diffuseness parameter= 0.60 fm,
$Y_{20} = \sqrt{\frac{5}{16\Pi}}(3\cos^2\theta - 1)$,
$Y_{30} = \sqrt{\frac{7}{16\Pi}}(5\cos^3\theta - 3\cos\theta)$, and
$Y_{40} = \frac{3}{16\sqrt{\Pi}}(35\cos^4\theta - 30\cos^2\theta + 3)$
are the spherical harmonics. The values of deformation parameters have been taken from the reference [17]. In our present work, we have studied 2 types of collision configuration of Pb nucleus which are chosen by the orientation of semi-major axis and semi-minor axis of the colliding deformed Pb nucleus. The details about the angular orientation in these two configurations in terms of $\theta$ and the direction of impact parameter are mentioned in Table 1. The z axis is the beam direction whereas x(b) represents the impact parameter direction along the x axis.

The two special geometrical configurations of collisions, body-body and tip-tip are controlled by $\theta$ while all the other coordinates are integrated over same range. However, as shown in article [37], by changing $\phi$, several other configurations can be made but we will focus on these two special ones. Since HYDJET++ model works in cylindrical polar coordinates $(\rho, \phi, z)$ whereas the nuclear density function is described in spherical polar coordinates $(r, \theta, \phi)$. Therefore, we perform a transformation of coordinates using the relations $\theta = \tan^{-1}(z/bc)$ and $\theta = \tan^{-1}(bc/z)$ for body-body and tip-tip geometrical configurations, respectively. The transformation equation $\rho = \sqrt{z^2 + r^2}$ is changed to $r = \sqrt{z^2 + bc^2}$. Thus, r becomes the new variable bc and $\rho$ is changed to r. The Woods-Saxon distribution with these parameters and orientations are sampled to obtain discrete nucleon positions and are fed as input into the Monte Carlo simulation. To validate the modification made in the deformed Woods-Saxon function and have a proper visualization, the nuclear density profiles in cylindrical coordinates for spherical lead, and deformed lead in body-type and tip-type configurations are shown in Fig. 1. Here, we see that upon deformation, nuclear density increases where tip-type is more than the body-type. The minimum bias nuclear density profile overlaps the profile for body-type configuration. This is because of the absence of quadrupole deformation parameter $\beta_2$. When a quadrupole deformation as high as 0.0544 [16] is introduced in the profile, the deformed nuclear density for body-type rises.

**Table 1**
Details of angular configuration in $^{208}$Pb-$^{208}$Pb collisions. t and p in subscript denotes the target and projectile, respectively.

| Parameter | Body-Body | Tip-Tip |
|---|---|---|
| $\theta_p$ | $\pi/2$ | 0 |
| $\theta_t$ | $\pi/2$ | 0 |
| Impact parameter direction | minor axis | minor axis |

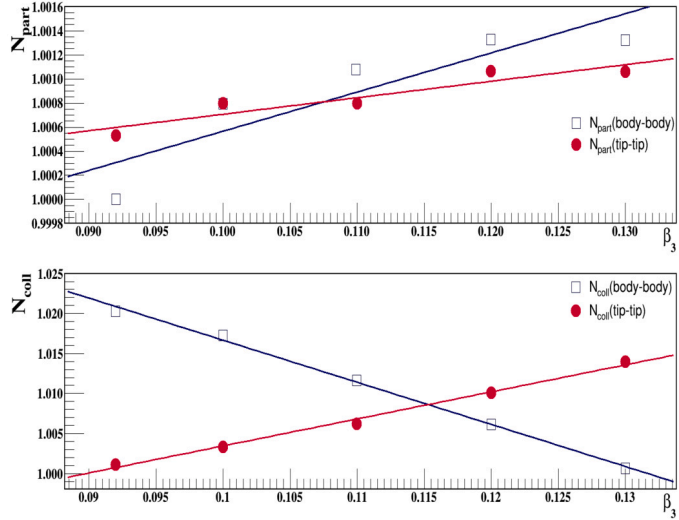

**Fig. 2.** Variation of number of participants $N_{part}$ and number of binary collisions $N_{coll}$ with respect to nuclear octupole deformation parameter $\beta_3$ in deformed Pb-Pb collisions at 5.02 TeV centre of mass energy.

To test the predictive capability of HYDJET++ model in body–body and tip–tip configurations of Pb-Pb collisions, the number of participants ($N_{part}$ in top panel) and number of binary collisions ($N_{coll}$ in bottom panel) in both the geometrical configurations are calculated and plotted in Fig. 2 with respect to nuclear octupole deformation parameter $\beta_3$. In the upper panel, it is seen that the number of participants for body-body collisions increase strongly (greater slope) with nuclear octupole deformation in comparison to tip-tip collisions (smaller slope). In the lower panel, the number of binary collisions are plotted against octupole deformation parameter where it is seen that the two geometrical configurations have opposite behaviour. Tip-tip collisions show a positive correlation while body-body collisions present a negative correlation of the number of binary collisions with the nuclear octupole deformation parameter $\beta_3$.

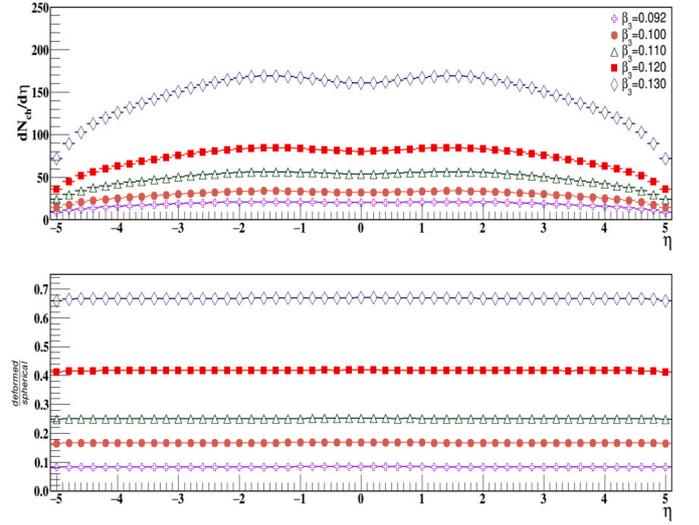

**Fig. 3.** Upper panel: Pseudorapidity distribution of primary charged particles in most-central (0–5)% class of Pb-Pb collisions at 5.02 TeV. The distribution for each case has been scaled by some value for proper visualization. Lower panel: The ratio of pseudorapidity distribution of deformed to spherical Pb-Pb collisions at 5.02 TeV. The figure shows results for five different values of (positive-) octupole deformation parameter.

We have produced $1.5 \times 10^5$ events for each most-central (0-5)% collisions with a certain value of $\beta_3$ and other orders of deformation assumed zero, for each geometrical configuration individually under the kinematic conditions $0 < p_T < 20$ GeV/c and $|\eta| < 0.8$ using HYDJET++ model. Some of the significant outcomes are discussed as follows:

The very first observable to be discussed is the pseudorapidity distribution of the primary charged hadrons (in Fig. 3). Such a observable is quite helpful in understanding the properties of the fireball formed in heavy-ion collisions and also the particle production mechanism involved. Another important reason here for the study here is to visualise the effect of nuclear octupole deformation on the particle production mechanism. We already know from recent works on deformed nucleus-nucleus collisions that, nuclear deformation affects (promotes or depletes) the particle production strength in deformed heavy-ion collisions ([18] and some references there within) but individual effect is quite left untouched. Although, the correlation between charged hadron multiplicity and initial geometrical configurations of nucleus-nucleus collisions is small [38], yet the effect is quite debative.

In Fig. 3, we present pseudorapidity distributions corresponding to five values of octupole deformation. The distribution for each case has been scaled by some value for proper visualization. Therefore, each distribution appears to have very small particle production. Here, it can be seen that the charged hadron pseudorapidity density increases as $\beta_3$ increases. The quantitative strength of the distribution strongly depends on the strength of octupole deformation. This is evident from the Fig. 3 where we observe that the charged hadron pseudorapidity density increases almost 30 times as octupole deformation was increased by a magnitude of 0.010 (0.120 to 0.130).

In the lower panel of Fig. 3, the ratio of pseudorapidity distribution for deformed Pb-Pb collisions to spherical Pb-Pb collisions is presented. The ratios are found to be less than 1. The ratio decreases by almost 33.5% for the maximum value of $\beta_3$. This fall in the particle production increases as $\beta_3$ decreases. This means that the deformed Pb-Pb collisions result in decreased production of particles. In Fig. 4, we present the pseudo rapidity distributions in body-body and tip-tip geometrical configurations. In most-central collisions, the charged hadron multiplicity does not differ by a huge quantity. Yet, we observe that tip-tip collisions are higher than body-body collisions by a difference of 10 to 30 similar to other deformed collision systems [18]. Qualitative behaviour is similar to that of the Fig. 3.

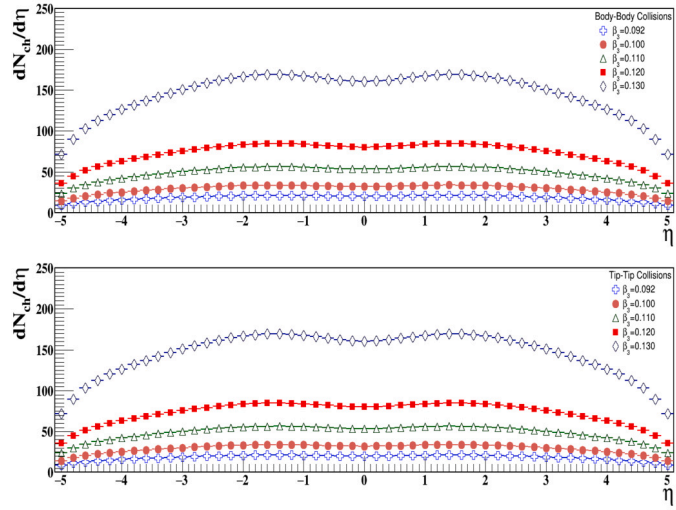

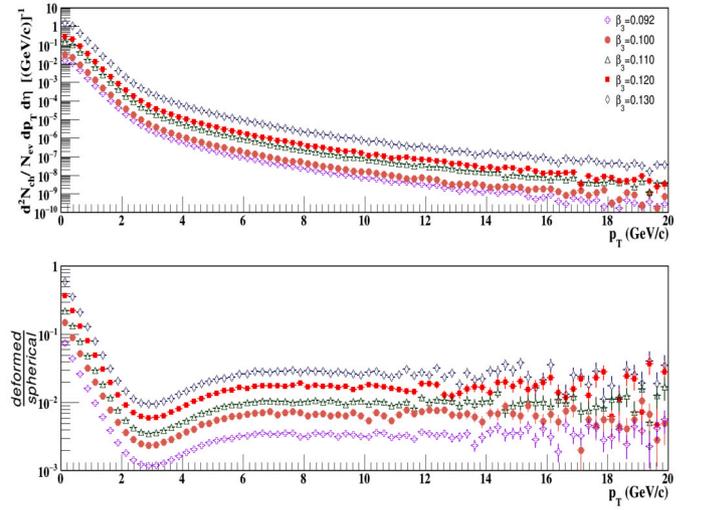

**Fig. 4.** Upper subfigure: Variation of $dN_{ch}/d\eta$ with respect to $\eta$ of primary charged hadrons in body-body configuration of Pb-Pb collisions over five different values of (positive-) octupole deformation parameter. Lower subfigure: Variation of $dN_{ch}/d\eta$ with respect to $\eta$ of primary charged hadrons in tip-tip configuration of Pb-Pb collisions over five different values of (positive-) octupole deformation parameter. The distribution for each case has been scaled by some value for proper visualization.

**Fig. 5.** Upper panel: Normalized transverse momentum distribution of primary charged particles in most-central (0–5)% class of Pb-Pb collisions at 5.02 TeV. The distribution for each case has been scaled by some value for proper visualization. Lower panel: The ratio of transverse momentum distribution of deformed to spherical Pb-Pb collisions at 5.02 TeV. The figure shows results for five different values of (positive-) octupole deformation parameter.

Another important observable to be analysed here is the transverse momentum distribution of primary charged hadrons. Fig. 5 represents the normalized minimum bias transverse momentum distribution of primary charged hadrons in deformed Pb-Pb collisions at 5.02 TeV. The distributions have been scaled by some factor to have a distinguishable picture. The figure comprises of two subfigures: the top subfigure presents the transverse momentum spectra as a function of $p_T$ for five values of octupole deformation parameter $\beta_3$, and the bottom subfigure depicts the ratios of deformed $p_T$ distributions to that of the spherical case. The transverse momentum distribution gradually decreases as $p_T$ increases. A positive correlation is observed between the $p_T$ spectra and nuclear octupole deformation parameter $\beta_3$. However, the quantitative strength varies with the strength of $\beta_3$. The difference can be more clearly visualized in the bottom subfigure. The ratios of deformed to spherical Pb-Pb collisions clearly show that, at low $p_T$, ($p_T < 3.0$ GeV/c), normalized $p_T$ distribution decreases with increasing $p_T$. Beyond this limit, the ratios increase and saturate or become independent of $p_T$. Another important observation is that upon being deformed, transverse momentum spectra decreases in Pb-Pb collisions.

In Fig. 6, we present normalized transverse momentum spectra with respect to $p_T$ for body-body and tip-tip collisions. In most-central collisions, the effect of initial state geometry is quite measurable. However, in our present work, we have only introduced the third order of deformation parameter, the octupole moment parameter, therefore the difference between body-body and tip-tip corresponding to a fixed value of $\beta_3$ is not expected to be very high. $p_T$ spectra is higher in tip-tip collisions than that in case of body-body collisions. In tip-tip collisions, the slope of the $p_T$ spectra is lower in comparison to body-body configuration. This is attributed to the effect of transverse or directed flow in the medium. Also, this indicates that the source temperature in body-body collisions is lower in comparison to tip-tip collisions. As a function of $\beta_3$, it is observed that the slope of $p_T$ distribution decreases as we introduce higher and higher value of octupole deformation $\beta_3$ in the nuclear density profile of Pb nucleus. This shows that the fireball temperature created in heavy-ion collision is higher in case of higher value of octupole deformation and decreases with decreasing $\beta_3$.

In Fig. 7, we present average anisotropic flow integrated over $p_T$ from 0.001 to 10 GeV/c as a function of $\beta_3$ in deformed Pb-Pb collisions at 5.02 TeV. It is observed that elliptic flow $v_2$ and triangular flow $v_3$ are dependent on nuclear octupole deformation parameter. Average elliptic flow is higher than average triangular flow. Average $v_2$ and $v_3$ flow in minimum bias deformed Pb-Pb collisions are almost independent of $\beta_3$. Flow for body-body collisions are higher than flow for tip-tip collisions. However, the nucleus-nucleus collisions in the two geometrical configurations show opposite behaviour. Body-body collisions show positive correlation (albeit weak) with $\beta_3$ while tip-tip collisions show a weak negative correlation with octupole deformation parameter $\beta_3$.

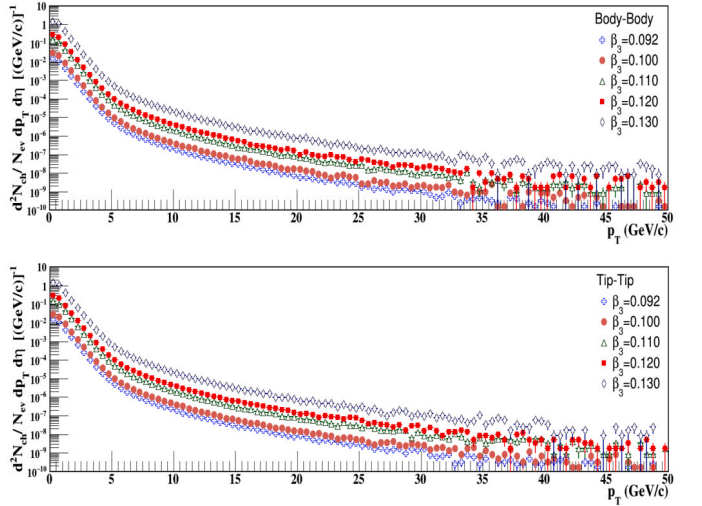

**Fig. 6.** Variation of normalized $p_T$-spectra with respect to transverse momentum $p_T$ of primary charged hadrons in body-body configuration (upper subfigure) and in tip-tip configuration (lower subfigure) of Pb-Pb collisions over five different values of (positive-) octupole deformation parameter. The distribution for each case has been scaled by some value for proper visualization.

By this time, a lot of study has been done to study QGP formation in deformed heavy-ion collisions such as U-U at RHIC and Xe-Xe at LHC energy regime. However, the individual role of the nuclear deformation parameters have not been quite explored from the domain of heavy-ion collisions. The source of these parameters are the nuclear structure studies performed at very low energies. The introduction of nuclear deformation alters the initial conditions in heavy-ion collisions. Thus, it is very important to study the individual effect of the deformation parameters. In this work, we incorporate only nonzero octupole deformation $\beta_3$ in nuclear density profile of $Pb^{208}$ and study the role or effect of this deformation on the particle production mechanism and other QGP observables by colliding this deformed Pb-Pb collisions in most-central region at LHC energy regime. In brief, we have tried to make a scrupulous study of Pb-Pb collisions at 5.02 TeV centre of mass energy. Performing this under the framework of HYDJET++ model offered us the possibility to study the system in various geometrical configurations which are cognizant of the initial conditions. In the present study, we have used body-body and tip-tip configurations for the analysis.

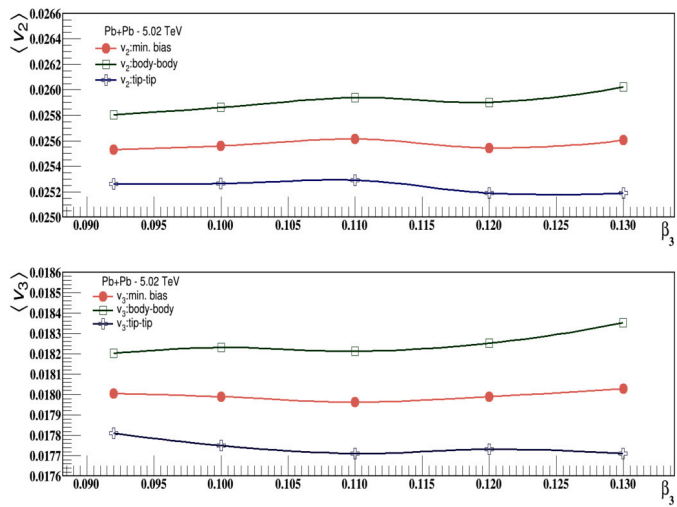

**Fig. 7.** Variation of average anisotropic flow $v_n$ (n=2,3) with respect to nuclear octupole deformation parameter $\beta_3$ in minimum bias deformed Pb-Pb collisions at 5.02 TeV centre of mass energy using HYDJET++ model. The figure presents comparison of the results with those from body-body and tip-tip collisions.

The charged hadron pseudorapidity density and transverse momentum spectra decrease when octupole deformation is being introduced in the nuclear density profile. These observables are found to be strongly dependent on the strength of the octupole deformation parameter. Tip-tip collisions have higher multiplicity and $p_T$ spectra than body-body collisions. A clear dependence of the particle multiplicity on the nuclear deformation and hence on the geometry of the colliding system is observed. Average elliptic flow $v_2$ and triangular flow $v_3$ show a weak dependence on the nuclear octupole deformation. Body-body collisions show a weak positive correlation while tip-tip collisions show a weak negative correlation with $\beta_3$. Thus, it is expected that our work will enlighten the role of octupole deformation in particle production mechanism at LHC energy regime. The effect of system size and geometry (through nuclear deformation via $\beta_3$) can be clearly visualized from our work. With these evidences, we can in future look deeper into the role of octupole deformation in the field of heavy-ion collisions.

BKS sincerely acknowledges financial support from the Institutions of Eminence (IoE) BHU grant number-6031. Saraswati Pandey acknowledges the financial support obtained from UGC and IoE under a research fellowship scheme during the work.

**Declaration of competing interest**

The authors declare the following financial interests/personal relationships which may be considered as potential competing interests:

Saraswati Pandey reports was provided by Banaras Hindu University. If there are other authors, they declare that they have no known competing financial interests or personal relationships that could have appeared to influence the work reported in this paper.

**Data availability**

Data will be made available on request.

**References**

[1] F. Karsch, in: 15th Int. Conf. on Ultra-Relativistic Nucleus-Nucleus Collisions, Quark Matter 2001, Nucl. Phys. A 698 (2002) 199, ISSN 0375-9474.
[2] Z.-W. Lin, C.M. Ko, B.-A. Li, B. Zhang, S. Pal, Phys. Rev. C 72 (2005) 064901.
[3] L. Ma, G.L. Ma, Y.G. Ma, Phys. Rev. C 94 (2016) 044915.
[4] B. Alver, G. Roland, Phys. Rev. C 81 (2010) 054905, https://doi.org/10.1103/PhysRevC.81.054905.
[5] Y.-G. Ma, S. Zhang, Influence of Nuclear Structure in Relativistic Heavy-Ion Collisions, Springer Nature Singapore, Singapore, ISBN 978-981-15-8818-1, 2020, pp. 1–30.
[6] P. Carzon, S. Rao, M. Luzum, M. Sievert, J. Noronha-Hostler, Phys. Rev. C 102 (2020) 054905.
[7] F.G. Gardim, F. Grassi, M. Luzum, J.-Y. Ollitrault, Phys. Rev. Lett. 109 (2012) 202302.
[8] P. Bożek, W. Broniowski, Phys. Rev. C 88 (2013) 014903.
[9] G. Giacalone, J. Noronha-Hostler, M. Luzum, J.-Y. Ollitrault, Phys. Rev. C 97 (2018) 034904.
[10] S. Acharya, F. Acosta, D. Adamová, J. Adolfsson, M. Aggarwal, G. Aglieri Rinella, M. Agnello, N. Agrawal, Z. Ahammed, S. Ahn, et al., Phys. Lett. B 784 (2018) 82, ISSN 0370-2693.
[11] S. Acharya, F.-. Acosta, D. Adamová, J. Adolfsson, M. Aggarwal, G. Aglieri Rinella, M. Agnello, N. Agrawal, Z. Ahammed, S. Ahn, et al., Phys. Lett. B 788 (2019) 166, ISSN 0370-2693.
[12] L. Adamczyk, J.K. Adkins, G. Agakishiev, M.M. Aggarwal, Z. Ahammed, I. Alekseev, J. Alford, A. Aparin, D. Arkhipkin, E.C. Aschenauer, et al., STAR Collaboration, Phys. Rev. Lett. 115 (2015) 222301.
[13] G. Giacalone, Phys. Rev. C 99 (2019) 024910.
[14] P. Möller, A. Sierk, T. Ichikawa, H. Sagawa, At. Data Nucl. Data Tables 109–110 (1) (2016), ISSN 0092-640X.
[15] L.M. Robledo, G.F. Bertsch, Phys. Rev. C 84 (2011) 054302.
[16] B. Pritychenko, M. Birch, B. Singh, M. Horoi, At. Data Nucl. Data Tables 107 (2016) 1, ISSN 0092-640X.
[17] T. Kibédi, R. Spear, At. Data Nucl. Data Tables 80 (2002) 35, ISSN 0092-640X.
[18] S. Pandey, S.K. Tiwari, B.K. Singh, Phys. Rev. C 103 (2021) 014903.
[19] S. Pandey, B.K. Singh, J. Phys. G, Nucl. Part. Phys. 49 (2022) 095001.
[20] S. Pandey, B.K. Singh, Eur. Phys. J. A 59 (2023).
[21] I. Lokhtin, L. Malinina, S. Petrushanko, A. Snigirev, I. Arsene, K. Tywoniuk, Comput. Phys. Commun. 180 (2009) 779, ISSN 0010-4655.
[22] I. Lokhtin, L. Malinina, S. Petrushanko, A. Snigirev, I. Arsene, K. Tywoniuk, arXiv preprint arXiv:0903.0525, 2009.
[23] I.P. Lokhtin, A.M. Snigirev, Eur. Phys. J. C Part. Fields 45 (2006) 211.
[24] J.D. Bjorken, Phys. Rev. D 27 (1983) 140.
[25] E. Braaten, M.H. Thoma, Phys. Rev. D 44 (1991) R2625.
[26] I.P. Lokhtin, A.M. Snigirev, Eur. Phys. J. C 16 (2000) 527, arXiv:hep-ph/0004176.
[27] R. Baier, Y.L. Dokshitzer, A.H. Mueller, D. Schiff, Phys. Rev. C 60 (1999) 064902.
[28] R. Baier, Y.L. Dokshitzer, A.H. Mueller, D. Schiff, Phys. Rev. C 64 (2001) 057902.
[29] B. Andersson, The Lund Model, vol. 7, Cambridge University Press, 2005.
[30] K. Tywoniuk, I. Arsene, L. Bravina, A. Kaidalov, E. Zabrodin, Phys. Lett. B 657 (2007) 170, ISSN 0370-2693.
[31] V. Gribov, Sov. Phys. JETP 29 (1969) 064905.
[32] B. Svetitsky, Phys. Rev. D 37 (1988) 2484.
[33] P.K. Srivastava, B.K. Patra, Eur. Phys. J. A 53 (2017) 116.
[34] N.S. Amelin, R. Lednicky, T.A. Pocheptsov, I.P. Lokhtin, L.V. Malinina, A.M. Snigirev, I.A. Karpenko, Y.M. Sinyukov, Phys. Rev. C 74 (2006) 064901.
[35] N.S. Amelin, R. Lednicky, I.P. Lokhtin, L.V. Malinina, A.M. Snigirev, I.A. Karpenko, Y.M. Sinyukov, I. Arsene, L. Bravina, Phys. Rev. C 77 (2008) 014903.
[36] S. Akkelin, P. Braun-Munzinger, Y. Sinyukov, Nucl. Phys. A 710 (2002) 439, ISSN 0375-9474.
[37] Md Rihan Haque, Zi-Wei Lin, Bedangadas Mohanty, Phys. Rev. C 85 (2012) 034905.
[38] A. Singh, P. Srivastava, O. Chaturvedi, S. Ahmad, B. Singh, Eur. Phys. J. C 78 (2018) 1.